\documentclass[preprint,showpacs]{revtex4}

\begin{document}

\title{MESOSCOPIC PHASE SEPARATION IN THE MODEL WITH COMPETING JAHN-TELLER AND
COULOMB INTERACTION.}

\begin{abstract}
We have derived the interaction between Jahn-Teller centers due to
optical and acoustic phonons. Without Coulomb repulsion the model
leads to a global phase separation at low temperature. At long
distances Coulomb repulsion always dominates the short range
attraction leading to phase separation on a short scale. On the
basis of Monte-Carlo simulation of microscopic lattice-gas model
we have formulated an effective phenomenological model. In the
limit of low density of polarons, phase separation takes place in
the form of charged bubbles. In the limit of high density, charge
segregation occurs on domain walls.
\end{abstract}

\date{\today}
\author{V.V. Kabanov, T. Mertelj, D. Mihailovic}
\address{J. Stefan Institute 1001, Ljubljana, Slovenia}

\maketitle

There is substantial evidence that the ground state in many of
oxides is inhomogeneous\cite{egami}. In cuprates, for example
neutron-scattering experiments suggest that phase segregation
takes place in the form of stripes or short segments of
stripes\cite{bianconi,mook}. There is some controversy whether
this phase segregation is associated with magnetic interactions.
On the other hand it is also generally accepted that the charge
density in cuprates is not homogeneous.

The idea of charge segregation in cuprates appeared just after
discovery of High-Tc materials \cite{zaanen,emery,gorkov}. In most
cases long-range Coulomb repulsion was not considered. Recently it
was suggested that interplay of the short range lattice attraction
and the long-range Coulomb repulsion between charge carriers could
lead to the formation  of short metallic \cite {fedia1,akpz} or
insulating\cite{akpz,fedia2} stripes of polarons. If the
attractive potential is isotropic, charged bubbles have a
spherical shape. Kugel and Khomskii \cite{klim} suggested recently
that the anisotropic attraction forces caused by Jahn-Teller
centers could lead to the phase segregation in the form of
stripes. The long-range anisotropic attraction forces appear as
the solution of the full set of elasticity equations (see
ref.\cite{eremin}). Alternative approach to take into account
elasticity potentials was proposed in ref.\cite{Shenoy} and is
based on the proper consideration of compatibility constraint
caused by absence of a dislocation in the solid.

Recently we formulated the model\cite{mk} where we suggested that
interaction of a two-fold degenerate electronic state with
$\tau_{1}$ phonon modes at a finite wave-vector can lead to a
local nonsymmetric deformation and short-length scale charge
segregation. It was suggested \cite{percolation} that this phase
separation may lead to the insulator to superconductor phase
transition governed by percolation. In this paper we reduce the
proposed model to the lattice gas model and show that the model
indeed displays phase separation, which may occur in the form of
stripes or clusters depending on anisotropy of the short range
attraction between localized carriers. We also generalize the
model taking into account interaction of the Jahn-Teller centers
via elasticity induced field.

We first derive a simplified lattice gas model. We show that the
model without Coulomb repulsion displays a first order phase
transition at a constant chemical potential. When the number of
particles is fixed, the system is unstable with respect to the
global phase separation below a certain critical temperature. In
the presence of the Coulomb repulsion, global phase separation
becomes unfavorable due to a large contribution to the energy from
long range Coulomb interaction. The system shows mesoscopic phase
separation where the size of charged regions is determined by the
competition between the energy gain due to ordering and energy
cost due to breaking of the local charge neutrality. Since the
short range attraction is anisotropic the phase separation may be
in the form of short segments or/and stripes.

Let us begin by considering a simplified version of the JT model
Hamiltonian \cite{mk}, taking only the deformation of the $B_{1g}$
symmetry:
\begin{equation}
H_{JT}=g\sum_{\mathbf{n},\mathbf{l}}\sigma _{3,\mathbf{l} }f(\mathbf{n}%
)(b_{\mathbf{l+n}}^{\dagger }+ b_{\mathbf{l+n}}),
\end{equation}
here the Pauli matrix $\sigma _{3,\mathbf{l}}$ describes two components of
the electronic doublet, and $f(\mathbf{n})=(n_{x}^{2}-n_{y}^{2})f_{0}(n)$ where $%
f_{0}(n)$ is a symmetric function describing the range of the interaction.
For simplicity we omit the spin index in the sum. The model could be easily
reduced to a lattice gas model. Let us introduce the classical variable $%
\Phi _{\mathbf{i} }=<b_{\mathbf{i}}^{+}+b_{\mathbf{i}}>/\sqrt{2}$
and minimize the energy as a function of $\Phi _{\mathbf{i}}$ in
the presence of the harmonic term $\omega \sum_{\mathbf{i}}\Phi
_{\mathbf{i}}^{2}/2$. We obtain the deformation, which corresponds
to the minimum of energy,
\begin{equation}
\Phi _{\mathbf{i}}^{(0)}=-\sqrt{2}g/\omega \sum_{\mathbf{n}}\sigma _{3,%
\mathbf{i}+\mathbf{n}}f(\mathbf{n}).
\end{equation}
Substituting $\Phi_{\mathbf{i}}^{(0)}$ into the Hamiltonian (1)
and taking into account that the carriers are charged we arrive at
the lattice gas model. We use a pseudospin operator $S=1$ to
describe the occupancies of the two electronic levels $n_{1}$and
$n_{2}$. Here $S^{z}=1$
corresponds to the state with $n_{1}=1$ , $n_{2}=0$, $S_{i}^{z}=-1$ to $%
n_{1}=0$, $n_{2}=1$ and $S_{i}^{z}=0$ to $n_{1}=n_{2}=0$.
Simultaneous occupancy of both levels is excluded due to a high
onsite Coulomb repulsion energy. The Hamiltonian in terms of the
pseudospin operator is given by
\begin{equation}
H_{LG}=\sum_{\mathbf{i},\mathbf{j}}(-V_{l}(\mathbf{i}-\mathbf{j})S_{\mathbf{%
i }}^{z}S_{\mathbf{j}}^{z}+V_{c}(\mathbf{i}-\mathbf{j})Q_{\mathbf{i}}Q_{%
\mathbf{j}}),
\end{equation}
where $Q_{\mathbf{i}}=(S_{\mathbf{i}}^{z})^{2}$. $V_{c}(\mathbf{n}%
)=e^{2}/\epsilon_{0}a(n_{x}^{2}+n_{y}^{2})^{1/2}$ is the Coulomb potential, $%
e$ is the charge of electron, $\epsilon _{0}$ is the static dielectric
constant and $a$ is the effective unit cell period. The anisotropic short
range attraction potential is given by $V_{l}(\mathbf{n})=g^{2}/\omega \sum_{%
\mathbf{m} }f(\mathbf{m})f( \mathbf{n}+\mathbf{m})$. The attraction in this
model is generated by the interaction of electrons with optical phonons. The
radius of the attraction force is determined by the radius of the
electron-phonon interaction and the dispersion of the optical phonons\cite
{akpz}.

A similar model can be formulated in the limit of continuous
media. In this case the deformation is characterized by the
components of the strain tensor. For the two dimensional case we
can define 3 components of the
strain tensor: $e_{1}=u_{xx}+u_{yy}$, $\epsilon =u_{xx}-u_{yy}$ and $%
e_{2}=u_{xy}$ transforming as the $A_{1g}$, $B_{1g}$ and $B_{2g}$
representations of the $D_{4h}$ group respectively. These
components of the tensor are coupled linearly with the two-fold
degenerate electronic state which transforms as the $E_{g}$ or
$E_{u}$ representation of the point group. Similarly to the case
of previously considered interaction with optical phonons we keep
the interaction with the deformation $\epsilon $ of the $B_{1g}$
symmetry only. The Hamiltonian without the Coulomb repulsion term
has the form:
\begin{equation}
H=g\sum_{\mathbf{i}}S_{\mathbf{i}}^{z}\epsilon _{\mathbf{i}}+\frac{1}{2}
\left( A_{1}e_{1,\mathbf{i}}^{2}+A_{2}\epsilon _{\mathbf{i}}^{2}+A_{3}e_{3,
\mathbf{i}}^{2}\right),
\end{equation}
where $A_{j}$ are corresponding components of the elastic modulus
tensor. The components of the strain tensor are not independent
\cite{Bishop,Shenoy} and satisfy to the compatibility condition.
The compatibility condition leads to a long range anisotropic
interaction between polarons. The Hamiltonian in the reciprocal
space has the form:
\begin{equation}
H=g\sum_{\mathbf{k}}S_{\mathbf{k}}^{z}\epsilon _{\mathbf{k}}+(A_{2}+A_{1}U(%
\mathbf{k}))\frac{\epsilon _{\mathbf{k}}^{2}}{2}.
\end{equation}
The Fourrier transform of the potential is given by:
\begin{equation}
U(\mathbf{k})=\frac{(k_{x}^{2}-k_{y}^{2})^{2}}{%
k^{4}+8(A_{1}/A_{3})k_{x}^{2}k_{y}^{2}}\text{.}
\end{equation}
By minimizing the energy with respect to $\epsilon_{k}$ and taking
into account the long-range Coulomb repulsion we again derive
Eq.(3). The anisotropic interaction potential
$V_{l}(\mathbf{n})=-\sum_{\mathbf{k}}\exp{%
(i\mathbf{k\cdot n})}\frac{g^{2}}{2(A_{2}+A_{1}U(\mathbf{k}))}$ is
determined by the interaction with the classical deformation and
is long-range. It decays as $1/r^{2}$ at large distances in 2D.
Since the attraction forces decay faster then the Coulomb
repulsion forces at large distances the attraction can overcome
the Coulomb repulsion at short distances, leading to a mesoscopic
phase separation.

Irrespective of whether the resulting interaction between polarons
is generated by acoustic or optical phonons the main physical
picture remains the same. In both cases there is an anisotropic
attraction between polarons on short distances. The interaction
could be either ferromagnetic or antiferromagnetic in terms of the
pseudospin operators depending on mutual
orientation of the orbitals. Without loosing generality we assume that $V(%
\mathbf{n})$ is nonzero only for the nearest neighbors.

Our aim is to study the model (Eq.(3)) at constant average
density,
\begin{equation}
n=\frac{1}{N}\sum_{\mathbf{i}}Q_{\mathbf{i}},  \label{eq_npart}
\end{equation}
where $N$is the total number of sites. However, to clarify the physical
picture it is more appropriate to perform calculations with a fixed chemical
potential first by adding the term $-\mu \sum_{\mathbf{i}}Q_{\mathbf{i}}$ to
the Hamiltonian (3).

Models such as (3), but in the absence of the long-range forces, were
studied many years ago on the basis of the molecular-field approximation in
the Bragg-Williams formalism \cite{lajz,siva}. The mean-field equations for
the particle density $n$ and the pseudospin magnetization $M=\frac{1}{N}\sum_{%
\mathbf{i}}S_{\mathbf{i}}^{z}$ have the form\cite{lajz}:
\begin{eqnarray}
M &=&\frac{2\sinh {(2zV_{l}M/{k}_{B}{T})}}{\exp {(-\mu /k}_{B}{T)}+2\cosh {%
(2zV_{l}M/{k}_{B}{T})}}  \label{eq_M} \\
n &=&\frac{2\cosh {(2zV_{l}M/{k}_{B}{T})}}{\exp {(-\mu /{k}_{B}T)}+2\cosh {%
(2zV_{l}M/{k}_{B}{T})}}
\end{eqnarray}
here $z=4$ is the number of the nearest neighbours for a square
lattice in 2D and ${{k}_{B}}${\ is the Boltzman constant}. A phase
transition to an ordered state with a finite $M$ may be of either
the first or the second order, depending on the value of $\mu $.
For $\mu
>0$ it is always of the second order and $n\rightarrow 1$ as
$T\rightarrow 0$. For the large negative values $\mu <-2zV_{1}$
the phase transition is absent and $n\rightarrow 0$ as
$T\rightarrow 0$. For the physically important case $-2zV_{1}<\mu
<0$ ordering occures as a result of the first order phase
transition. Two solutions of Eqs.(8,9) with $M=0$ and with finite
$M$ corresponds to two different minima of the free energy. The
line of the phase transition is determined by the condition:
$F(M=0,\mu ,T)=F(M,\mu ,T)$ where $M$ is the solution of Eq. (8)
(Fig.1)). When the number of particles is fixed (Eq.9)
the system is unstable with respect to global phase separation below $%
T_{crit}$. As a result at fixed average $n$ two phases with $n_{0}=n(M=0,\mu
,T)$ and $n_{M}=n(M,\mu ,T)$ coexist.

To investigate further the properties of the system, we performed
Monte-Carlo (MC) simulations of the Hamiltonian Eq.(3). The simulations were
performed on a square lattice with dimensions up to $L\times L$ sites with $%
10\leq L\leq 100$ using a standard Metropolis algorithm\cite{Metropolis53}
in combination with simulated annealing\cite{Kirkpatrick83}.

First, for comparison with MF theory, the Monte-Carlo simulation of the
model Eq.(3) in absence of the Coulomb forces shows the reduction of $%
T_{crit}$ due to fluctuations in 2D, by a factor of $\sim 2$ (Fig.1).

Next we include the Coulomb interaction $V_{c}(r)$. We use open boundary
conditions to avoid complications due to the long range Coulomb forces and
ensure global electroneutrality by adding an electrostatic potential due to
the uniformly charged background $V_{jell}\left( \mathbf{i}\right)$ \ to
(3), such that the Monte Carlo interaction becomes $H_{MC}=H_{LG}+\sum_{%
\mathbf{i}}Q_{\mathbf{i}}V_{jell}\left( \mathbf{i}\right)$, with the total
background charge being equal in magnitude to the total charge of carriers, $%
e\sum_{\mathbf{i}}Q_{\mathbf{i}}$, but with opposite sign. The dimensionless
temperature $t=$ $k_{B}T\epsilon _{0}a/e^{2}$, and the energy per particle
are defined as $e_{MC}=H_{MC}\epsilon _{0}a/Ne^{2}$. The short range
potential $v_{l}(\mathbf{i})=V_{l}(\mathbf{i})\epsilon_{0}a/e^{2}$ was taken
to be nonzero only for $\left\vert \mathbf{i}\right\vert <2$ and was
therefore specified for nearest neighbours and next nearest neighbours as $%
v_{l}(1,0)$ and $v_{l}(1,1)$ respectively.

The results of the Monte Carlo simulations could be summarized as
follows. ($i$) Coulomb repulsion leads to further reduction of the
onset temperature of phase separation (see Fig.1). ($ii$)
Depending on the value of the attractive potential and its
anisotropy phase separation takes place in the
form of bubbles, horizontal and vertical or diagonal stripes (Fig.2). ($iii$%
) Depending on the anisotropy of the potential stripes (bubbles) are
ferromagnetically or antiferromagnetically orbitally ordered (Fig.2). ($iv$)
The size of clusters is determined by the ratio of short range attraction
and long-range Coulomb energy.

The results of the Monte-Carlo simulation of the model (3) allow general
model independent interpretation. Let us consider the classical free energy
density corresponding to the first order phase transition:
\begin{equation}
F_{1}=((t-1)+(\Lambda ^{2}-1)^{2})\Lambda ^{2}
\end{equation}
Here $t$ is dimensionless temperature. Let us assume that the order
parameter $\Lambda $ is coupled to the local charge density $\rho $, $%
F_{coupl}=\alpha (\Lambda ^{2}-\rho )^{2}$. The total free energy density
should contain the gradient term $F_{grad}=C(\nabla \Lambda )^{2}$ and
electrostatic energy $F_{el}=K\phi \rho $ as well. The electrostatic
potential is determined from the Poisson equation $\epsilon _{0}\Delta \phi
=4\pi e(\rho -\rho _{0})$. Here we write $\rho _{0}$ explicitly to take into
account global electroneutrality. Substituting the solution of the Poisson
equation to the electrostatic energy we obtain $F_{el}=K^{^{\prime }}\sum_{%
\mathbf{k}}V(k)(\rho -\rho _{0})_{\mathbf{k}}^{2}$, here $V(k)=1/k$.
Minimization of $F_{1}+F_{coupl}+F_{grad}+F_{el}$ at fixed $t$ and $\rho
_{0} $ gives the spatial variation of the order parameter. The results of
minimization are presented in Fig.3. As clearly seen from Fig.3 at low
density phase separation takes place in the form of bubbles of optimal
radius which are ordered. When density increases charge segregation appears
in the form of charged domain walls.

Let us assume that a single bubble of the ordered phase with the
radius $R$ has appeared. As it was discussed in \cite{mk,gorkov2}
the energy of the bubble is determined as: $\epsilon =-F\pi
R^{2}+\alpha \pi R+\gamma \pi R^{3}$,
here $F$ is the energy difference between two minima in the free energy, $%
\alpha $ is the surface energy, and $\gamma$ determines the energy due local
breaking of the charge neutrality. If $\alpha <\pi F/3\gamma $ $\epsilon $
has well defined minimum. This minimum corresponds to optimal size of the
cluster. Residual interactions leads to interbubble interactions and
ordering of clusters at low temperatures.

We have demonstrated that anisotropic interaction between
Jahn-Teller centers generated by optical and/or acoustical phonons
leads to the short scale phase separation in the presence of the
long range Coulomb repulsion. Topology of texturing differs from
charged bobbles to oriented charged stripes depending on the
anisotropy of short range potential.

\bigskip

\section{Figure Captions}

Fig. 1 The phase diagram of the model in absence of the Coulomb
repulsion. The full symbols represent the mean field (MF) solution
while empty symbols represent the Monte-Carlo (MC) simulation
result on systems with two different sizes $L$.

Fig. 2 Snapshots of clusters ordering at $t=0.04$, $n=0.2$ and $%
v_{l}(1,0)=-1 $ as a function of $v_{l}(1,1)$.

Fig. 3 Snapshots of the order parameter distribution at different average
densities obtained by minimization of the free energy (10). With increasing
of density crossover from charged bubbles to the charged domain walls is
clearly observed.

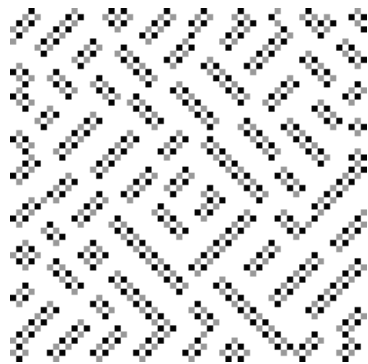
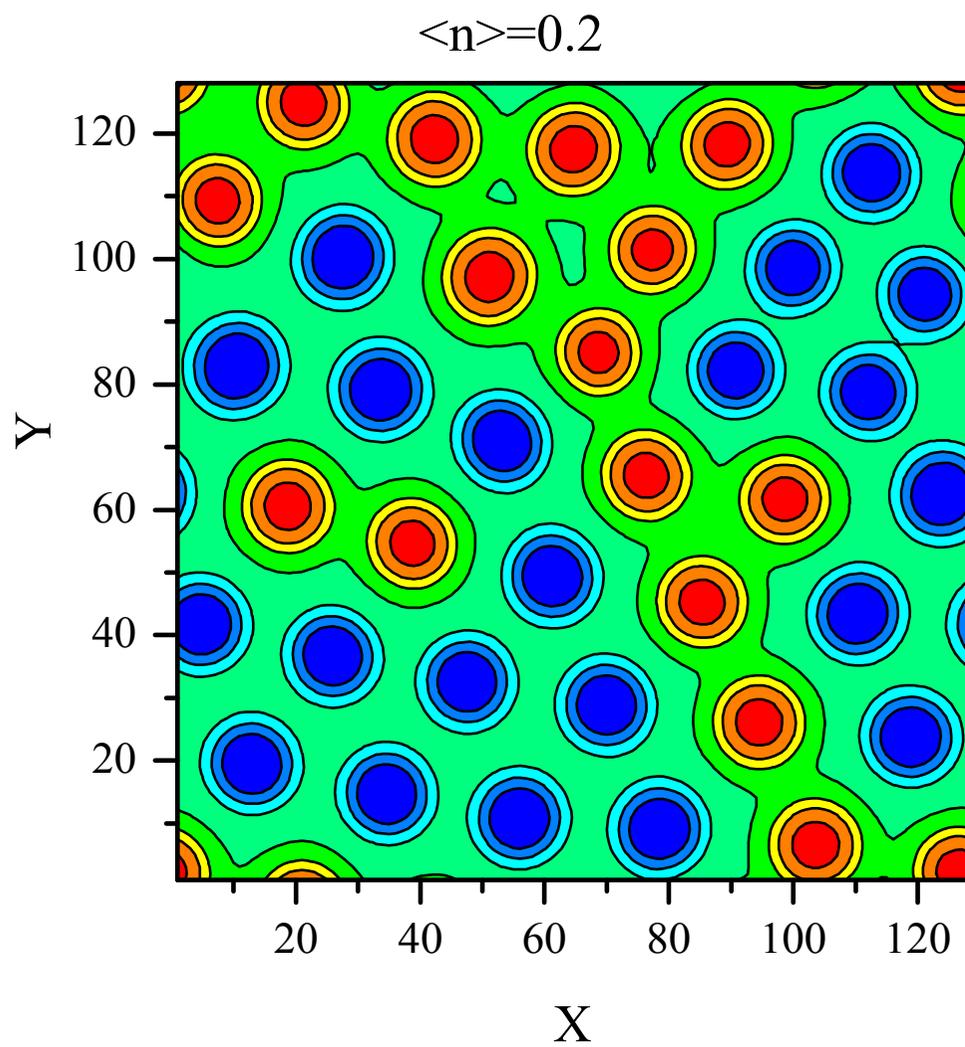
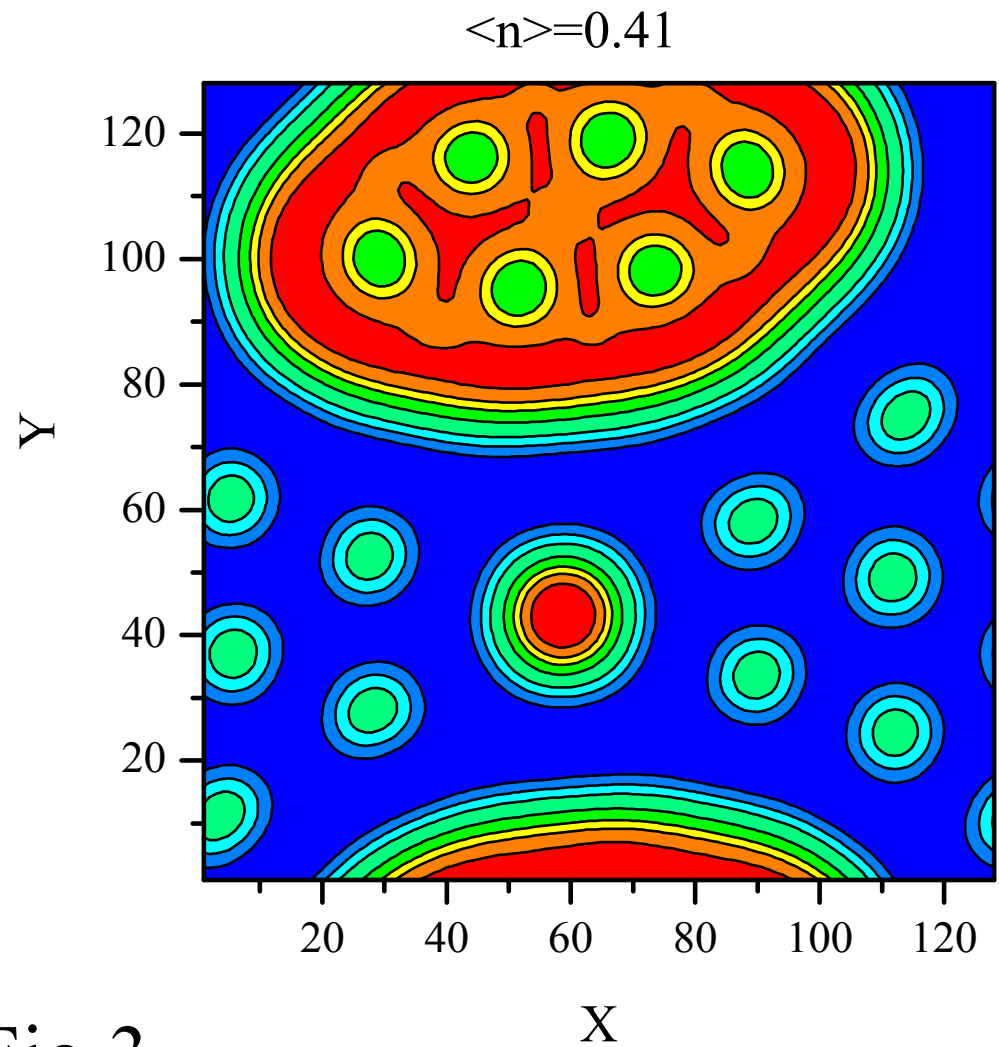
\end{document}